\documentclass{ws-mplaXXX}

\begin{document}
\markboth{W.~Detmold, W.~Melnitchouk \& A.~W.~Thomas}
{Extraction of parton distributions from lattice QCD}

\title{Extraction of parton distributions from lattice QCD}

\author{W.~Detmold}
\address{Department of Physics, University of
  Washington, Box 351560, Seattle, WA 98195, U.S.A.}

\author{W.~Melnitchouk}
\address{Jefferson Lab, 12000 Jefferson
  Avenue, Newport News, VA 23606, U.S.A.}

\author{A.~W.~Thomas}
\address{
Special Research Centre for the Subatomic Structure of Matter, \\
and Department of Physics and Mathematical Physics, \\
  University of Adelaide, Adelaide, SA 5005, Australia.}

\maketitle


\begin{abstract}
  We review the calculation of moments of both the polarized and
  unpolarized parton distribution functions of the nucleon in lattice
  QCD, and in particular their extrapolation to the physical region.
  We also discuss the reconstruction of the $x$ dependence of the
  valence quark distributions in the nucleon from a finite number of
  lattice moments.
%
\end{abstract}


\section{Introduction}
\label{sec:introduction}

One of the defining features of any hadron are its quark and gluon (or
generically, parton) momentum distributions.  Considerable information
has been accumulated on parton distribution functions (PDFs) of the
nucleon and nuclei from deep inelastic scattering and other high
energy experiments.  Information from the Drell-Yan reaction in $\pi
N$ scattering has also been used to determine PDFs of the pion.

Theoretically, the operator product expansion (OPE) in QCD allows one
to isolate the soft, non-perturbative PDFs from the hard,
perturbatively calculable scattering processes.  This factorization
property gives rise to a universal set of PDFs which can be used to
describe various reactions.  Because PDFs parameterize the
interactions of quarks and gluons in the hadron over all distance
scales, they are sensitive to the dynamics of QCD in the strong
coupling regime -- {\em i.e.}, at scales $Q^2 \sim \Lambda_{\rm
  QCD}^2$.  The only rigorous theoretical approach currently capable
of calculating observables in the non-perturbative regime from first
principles is lattice QCD.

The first exploratory studies of PDFs on the lattice were made almost
two decades ago.\cite{Mar87} Advances in computing power have recently
enabled dedicated simulations to be undertaken which for the first
time can be directly compared with
phenomenology.\cite{Dol02,QCDSF,KEK,Sas02} Since the numerical
simulations are performed in Euclidean space-time, it is not possible
to compute the PDFs directly as a function of the light-cone momentum
fraction, $x$ (PDFs are formally defined as light-cone correlation
functions involving currents with space-time separation $z^2 - (ct)^2
\approx 0$).  On the other hand, using the OPE one can formally
express the {\em moments} of PDFs in terms of matrix elements of local
operators between hadron states. These matrix elements can be
calculated numerically on the lattice.

Despite the impressive recent progress in the numerical simulations,
several approximations are still necessary in order to relate the
lattice moments to experiment.  Firstly, since space-time is
discretized on the lattice, with some finite lattice spacing $a$, the
results must be extrapolated to the continuum limit, $a \to 0$.  The
finite number of lattice sites also means that an extrapolation to the
infinite volume limit is necessary to avoid omitting important physics
arising from the long-range part of the nucleon wave function.  In
addition, since the computational cost of simulations scales with the
quark mass roughly as $m_q^{-4}$, current lattice simulations are
performed with unphysically large masses for $u$ and $d$ quarks,
typically $m_q^{\rm latt} > 30$~MeV, so that an extrapolation to
physical masses, $m_q^{\rm phys} \approx 5$~MeV, is
essential.\cite{Thomas:2002sj}

The chiral ($m_q$) extrapolation is found to play a particularly vital
role in understanding the connection between the lattice results and
phenomenology.  Whereas the lattice simulations yield
results\cite{Dol02,QCDSF} for the moments of unpolarized quark
distributions in the nucleon which are typically 50\% larger than
experiment when extrapolated linearly to $m_q^{\rm phys}$, inclusion
of the non-linear dependence on $m_q$ arising from the long-range
structure of the nucleon removes most of the discrepancy.\cite{Det01}
Furthermore, significant finite volume effects have been
found\cite{Sas02} in the calculation of the axial vector charge $g_A$,
which may eventually explain the residual discrepancy between the
extrapolated lattice value and experiment.

In this mini-review we focus on the challenges involved with
extrapolating the lattice data from the currently accessible regions
of parameter space to the physical region.  Using constraints
available from chiral effective theory, in Sec.~2 we review
extrapolation formulas which seek to describe the lattice data over a
large range of quark masses, and in the chiral limit.
The moments of both the spin-averaged and spin-dependent quark
distributions in the nucleon are considered (a corresponding analysis
of the moments of the quark distributions in the pion was also
undertaken recently\cite{Det03pi}).
The more ambitious task of reconstructing
the $x$ dependence of PDFs from a finite number of moments is
discussed in Sec.~3.  Because of the small number of lattice moments
available, only valence distributions can be analyzed at present; sea
quark distributions await future lattice data on higher moments of
PDFs which are necessary for an independent determination of the
valence and sea components.  Finally, in Sec.~4 we draw some
conclusions, and outline prospects for the reconstruction of PDFs from
future lattice simulations.

\newpage

\section{Extracting physical results from lattice simulations}
\label{sec:extr-phys-results}

The moments of parton distributions in the nucleon are formally
defined as
\begin{eqnarray}
  \langle x^n \rangle_{q}&=&\int_0^1 dx\,
  x^n\ [q(x)-(-1)^n \bar q(x)]\ ,
\label{mom:unp}                                 \\
  \langle x^n \rangle_{\Delta q} &=&\int_0^1 dx\,
  x^n\ [\Delta q(x)+(-1)^n \Delta \bar  q(x)]\ ,
\label{mom:hel}                                 \\
  \langle x^n \rangle_{\delta q} &=&\int_0^1 dx\,
  x^n\ [\delta q(x)-(-1)^n \delta \bar q(x)] \,,
\label{mom:trans}
\end{eqnarray}
for the spin-averaged, helicity and transversity distributions,
respectively.  From their definition, the moments alternate between
the total ($q+\overline q$) and valence ($q-\overline q$)
distributions, depending on whether $n$ is even or odd.  Using the
OPE, these moments can be related to ground state hadron matrix
elements of specific twist-two operators,\cite{Gockeler:1996mu} which
are calculated on the lattice.

In this review we focus not on the details of the lattice simulations
of the moments (a comprehensive survey of results was recently given
by the authors\cite{Det02ga}), but on their physical interpretation.
In particular, we examine the consequences of chiral symmetry for the
extrapolation of PDF moments to physical values of the quark mass,
$m_q$.

\subsection{Chiral symmetry}
\label{sec:chiral-symmetry}

The importance of chiral symmetry and the role of the pion cloud in
hadronic physics is well known.\cite{Thomas:1982kv,Detmold:2001hq} At
small quark masses, hadronic observables can be systematically
expanded in a series in $m_q$ within the framework of chiral
perturbation theory ($\chi$PT).  While the expansion coefficients are
generally free parameters, one of the unique consequences of
spontaneous chiral symmetry breaking in QCD is the appearance of terms
involving odd powers or logarithms of $m_\pi$.  From the
Gell-Mann--Oakes--Renner relation, which relates the quark and pion
masses at small $m_\pi$, $m_\pi^2 \sim m_q$, one finds that such terms
are non-analytic in the quark mass.  Furthermore, their coefficients,
which are determined from the infrared behavior of the pion loops, are
generally model independent.

The non-analytic term involving the lowest power of $m_\pi$ is known
as the ``leading non-analytic'' (LNA) term in the expansion.  For the
moments of the nucleon PDFs this was shown by Thomas {\em et
  al.}\cite{Tho00} to have the generic behavior $m_\pi^2 \log m_\pi$
arising from $\pi N$ intermediate states.  This was later confirmed in
$\chi$PT, where the coefficients of these terms were also calculated,
both for the nucleon,\cite{Arn02,Che01} and for the pion.\cite{Arn02}
Using these constraints, a low order chiral expansion for the moments
of the PDFs was developed recently by Detmold {\em et
  al.},\cite{Det01,Det03pi,Det02ga,Det01epj,Det02ijmp} which
incorporated the LNA behavior of the moments as a function of $m_q$
and also provided a connection to the heavy quark limit.

\subsection{Chiral extrapolation of PDF moments}
\label{sec:chir-extr-pdf}

Direct comparison of the currently available lattice moments of
nucleon PDFs with phenomenological ones is non-trivial.  The lattice
moments in general receive contributions from diagrams in which the
local operator insertions are on quark lines which are connected to
the nucleon source, as well as those where the operator acts on a
quark loop in the vacuum, which is connected only by gluon lines to
the quark lines originating in the nucleon source.
Obtaining a signal from the latter, so-called ``disconnected graphs''
is extremely difficult on the lattice, as recent
studies\cite{Lewis:2002ix} of the strangeness form factors of the
nucleon have shown, and to date only the connected contributions to
the lattice PDF moments have been computed.
Fortunately, the near degeneracy of the $u$ and $d$ quark masses, which
makes charge symmetry such a good symmetry of nature, means that the
disconnected diagrams cancel very precisely in the isovector
difference between the $u$ and $d$ PDFs of the proton.
An unambiguous comparison of the connected isovector PDF lattice moments
can therefore be made with the analogous experimental moments.

The mass dependence of the moments of the unpolarized isovector PDF of
the nucleon can be parameterized as\cite{Det01epj}
\begin{equation}
\label{extrap}
       \langle x^n\rangle_{u-d}\
        =\ a_n  \left( 1 + c_{\rm LNA} m_\pi^2
                          \log\frac{m_\pi^2}{m_\pi^2+\mu^2}
                \right)\
        +\ b_n \frac{m_\pi^2}{m_\pi^2+m_{b,n}^2}\ ,
\end{equation}
where (for $n>0$) the chiral coefficient is given by\cite{Arn02,Che01}
$c_{\rm LNA} = -(1 + 3 g_A^2)/(4\pi f_\pi)^2$.  Although we shall work
only with full QCD, we note that the corresponding coefficients have
also been computed in quenched and partially-quenched chiral
perturbation theory.\cite{PQXPT}
The $n=0$ moment, which corresponds to isospin charge, is not
renormalized by pion loops.  The argument of the logarithm corresponds
to the case where the pion loops are regulated with a simple, sharp
three-momentum cut-off.  The parameter $\mu$ has the effect of
suppressing the rapid variation of the chiral logarithm away from the
chiral limit.  Physically it is related to the size of the nucleon
core, which acts as the source of the pion field.\cite{Detmold:2001hq}
Since the isovector distribution $u(x)-d(x) \to \delta(x-1/3)$ in the
heavy quark limit ({\em i.e.}, as $m_q \to \infty$), one may choose to
constrain the constant $b_n$ in Eq.~(\ref{extrap}) by
\begin{equation}
b_n = \frac{1}{3^n} - a_n \left( 1 - \mu^2 c_{\rm LNA} \right)\ .
\end{equation}

\begin{figure}
\begin{center}
 \includegraphics[width=\columnwidth]{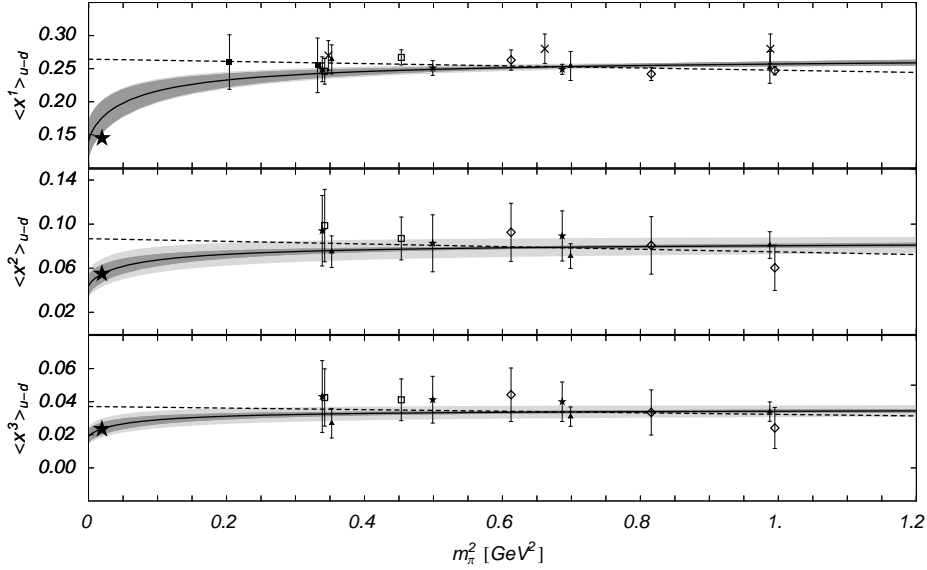}
\end{center}
\caption{\label{fig:unpolmom}
  Moments of the unpolarized $u-d$ distribution in the proton, for
  $n=1$, 2 and 3.  Lattice data\protect\cite{Det02ga} include both
  quenched (solid symbols) and unquenched (open symbols) results.  The
  solid line represents the full chiral extrapolation, while the inner
  (darkly shaded) error band shows variation of $\mu$ by $\pm$ 20\%,
  with the outer band (lightly shaded) showing the additional effects
  of shifting the lattice data within the extent of their error bars.
  Linear extrapolations are indicated by dashed lines, and the
  phenomenological values\protect\cite{UNPOLPARAM} are shown as
  large stars at the physical pion mass.}
\end{figure}

The results of the fits to the lattice data for the $n=1$, 2 and 3
moments of the unpolarized $u-d$ distribution using Eq.~(\ref{extrap})
are shown in Fig.~1.  In each plot the central curve is shown with two
error bands: the inner (darkly shaded) band shows variation of $\mu$
by $\pm$ 20\%, whilst the outer (lightly shaded) shows the additional
effects that result from shifting the lattice data up and down within
the extent of their error bars.  The extrapolated moments at the
physical pion mass are listed in Table~\ref{tab:restab}.
The fits to the data are quite insensitive to the choice of $m_{b,n}$
(as long as it is large),\cite{Det01epj} and it has been set to 5~GeV
for all $n$.  In practice, fits to the lattice data in which the heavy
quark limit is not guaranteed ({\em i.e.}, where the term involving
$b_n$ is simply $b_n m_\pi^2$) and $b_n$ is a third fitting
parameter,\cite{Det01} are indistinguishable from those in Fig.~1.

Note that the majority of the data points (filled symbols) are
obtained from simulations employing the quenched approximation (in
which background quark loops are neglected) whereas Eq.~(\ref{extrap})
is based on full QCD with quark loop effects included.  On the other
hand, recent calculations with dynamical quarks suggest that at the
relatively large pion masses ($m_\pi > 0.5$--0.6~GeV) where the full
simulations are currently performed, the effects of quark loops are
largely suppressed, as the data in Fig.~1 (small open symbols)
indicate.
Further details of the lattice data,\cite{Dol02,QCDSF,KEK,Sas02} and a
more extensive discussion of the fit parameters, can be found
elsewhere.\cite{Det02ga}

\begin{table}[h]
\tbl{Experimental moments\protect\cite{UNPOLPARAM,POLPARAM} and
	moments extrapolated\protect\cite{Det02ga} from lattice
	data.
\label{tab:restab}}
{\begin{tabular}{|c|cc|}                         
\hline \hline
Moment  & Experiment	& Extrapolated	\\ \hline \hline
$\langle x^1 \rangle_{u-d}$
        & 0.145(4)      & 0.18(3)       \\
$\langle x^2 \rangle_{u-d}$
        & 0.054(1)      & 0.05(2)       \\
$\langle x^3 \rangle_{u-d}$
        & 0.022(1)      & 0.02(1)       \\
$\langle x^0 \rangle_{\Delta u-\Delta d}$
        & 1.267(4)      & 1.12(6)       \\
$\langle x^1\rangle_{\Delta u-\Delta d}$
        & 0.210(25)     & 0.27(2)       \\
$\langle x^2 \rangle_{\Delta u-\Delta d}$
        & 0.070(11)     & 0.14(4)       \\
$\langle 1 \rangle_{\delta u-\delta d}$
        & ---           & 1.23(7)       \\
$\langle x \rangle_{\delta u-\delta d}$
        & ---           & 0.51(9)        \\ \hline\hline
\end{tabular}}
\end{table}

A similar analysis leads to analogous lowest order LNA
parameterizations of the mass dependence of the spin-dependent
moments\cite{Det02ijmp}
\begin{equation}
  \label{pxtrap}
  \langle x^n\rangle_{\Delta u-\Delta d}\ = \Delta a_n \left( 1 +
  \Delta c_{\rm LNA}m_\pi^2\log\frac{m_\pi^2}{m_\pi^2+\mu^2} \right)\
  +\ \Delta b_n\frac{m_\pi^2}{m_\pi^2+m_{b,n}^2}\,,
\end{equation}
and
\begin{equation}
  \label{txtrap}
  \langle x^n\rangle_{\delta u-\delta d}\ = \delta a_n \left( 1 +
  \delta c_{\rm LNA}m_\pi^2\log\frac{m_\pi^2}{m_\pi^2+\mu^2} \right)\
  +\ \delta b_n\frac{m_\pi^2}{m_\pi^2+m_{b,n}^2}\,,
\end{equation}
where the LNA coefficients are given by\cite{Che01}
$\Delta c_{\rm LNA}=-(1 + 2 g_A^2)/(4\pi f_\pi)^2$
and
$\delta c_{\rm LNA}=-(1 + 4 g_A^2)/2(4\pi f_\pi)^2$.
In the heavy quark limit, where spin-flavor symmetry is exact,
both $\Delta u(x)-\Delta d(x)$ and $\delta u(x)-\delta d(x)$ are
given by\cite{Gockeler:1996bm} $\frac{5}{3}\delta(x-1/3)$,
which leads to the constraints
\begin{equation}
  \Delta b_n=\frac{5}{3^{n+1}} - \Delta a_n \left( 1 - \mu^2 \Delta
    c_{\rm LNA} \right)\,,
\end{equation}
and
\begin{equation}
  \delta b_n=\frac{5}{3^{n+1}} - \delta a_n
  \left( 1 - \mu^2 \delta c_{\rm LNA} \right)\, .
\end{equation}
These are the most general lowest order parameterizations of the
twist-2 PDF moments consistent with chiral symmetry and the heavy
quark limits of QCD.

The $n=0$ moment of the spin-dependent isovector PDF is equivalent to
the axial charge of the nucleon, $g_A$.  It is well
known,\cite{Thomas:kw} for instance through the Adler-Weisberger sum
rule, that the $\Delta$ resonance plays a key role in $g_A$.  In the
framework of the chiral expansion, vertex renormalization
contributions involving a $\Delta$ isobar, although not leading
non-analytic, are large and extremely important in countering the
effect of wave function renormalization.\cite{Det02ga,HEMMERT} Indeed,
it is entirely because of the explicit role of the $\Delta$ that the
bare and renormalized $\pi NN$ couplings are
close.\cite{Theberge:1981mq}

\begin{figure}
\begin{center}
  \includegraphics[width=\columnwidth]{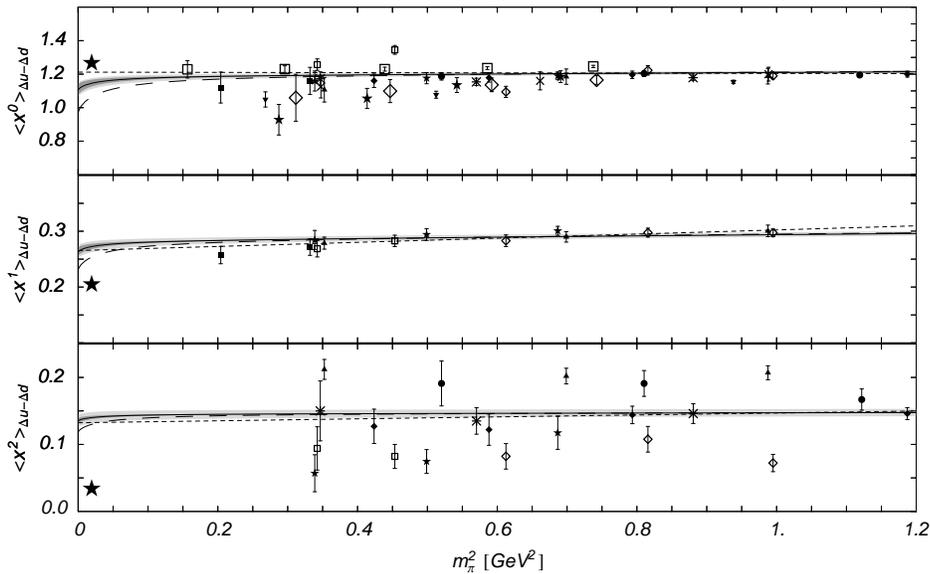}
\end{center}
\caption{\label{fig:polmom}
  Moments of the helicity distribution $\Delta u - \Delta d$ in the
  nucleon for $n=0$, 1 and 2.  The long-dashed curves show
  extrapolations neglecting the $\Delta$.
  The meaning of all other symbols is as in Fig.~1.}
\end{figure}

\begin{figure}
\begin{center}
 \includegraphics[width=\columnwidth]{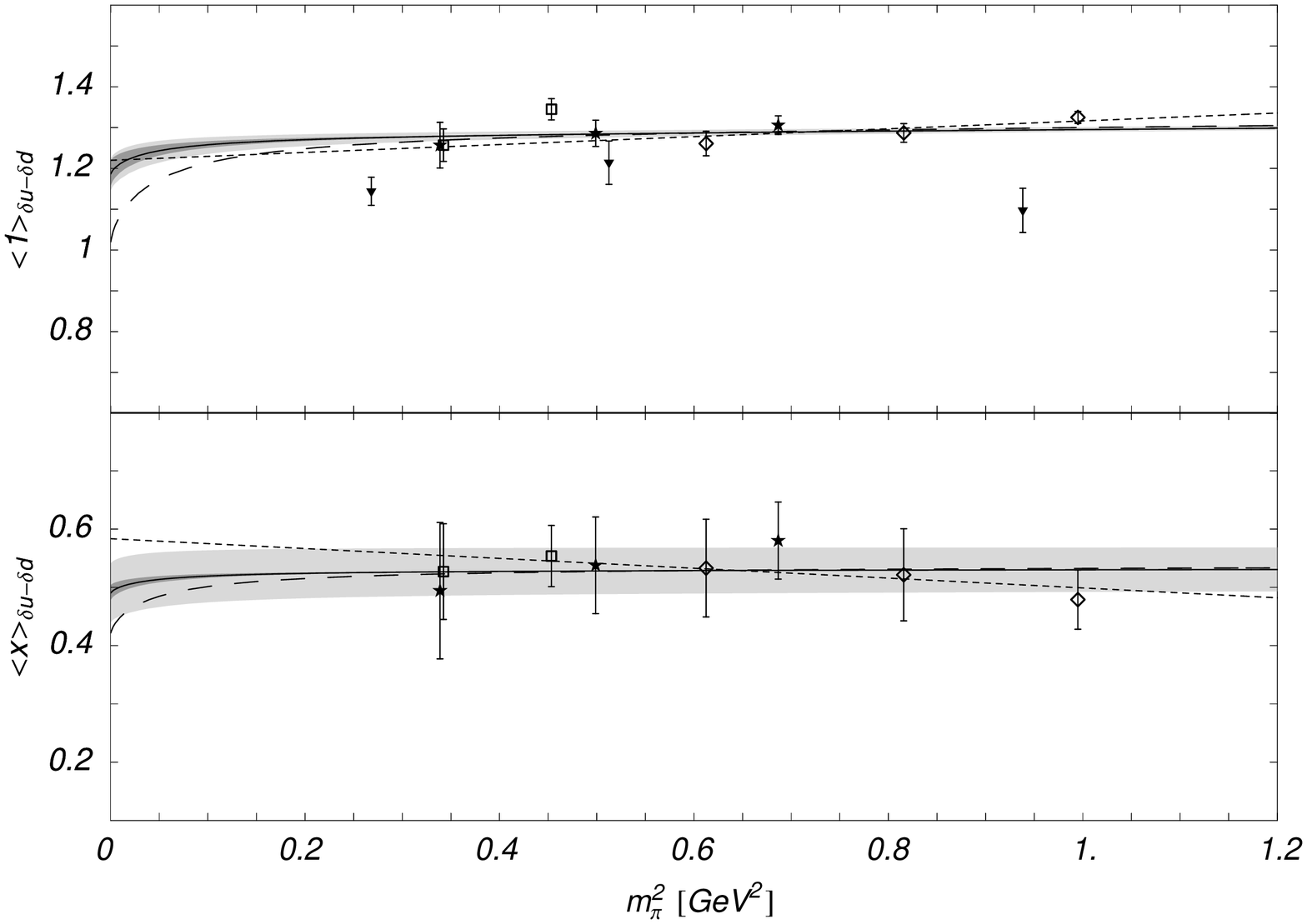}
\end{center}
\caption{\label{fig:transmom}
  Moments of the isovector transversity distribution
  $\delta u - \delta d$ for $n=0$ and 1.
  Symbols and curves are as in Fig.~2.}
\end{figure}

From the point of view of the chiral extrapolation problem, the
tendency for cancellation between wave function and polarized vertex
corrections means that, once virtual $\pi\Delta$ loops are
included, the chiral variation as $m_\pi \rightarrow 0$ is
approximately linear.  That is, for the polarized moments the chiral
non-analytic behavior produces far less curvature as one approaches
the physical pion mass than in the unpolarized case.  This can be seen
explicitly in Figs.~2 and 3, where the $n=0$, 1 and 2 moments of the
helicity distribution $\Delta u(x)-\Delta d(x)$ and the $n=0$ and 1
moments of the transversity distribution $\delta u(x)-\delta d(x)$ are
plotted, respectively.  In these figures, the long-dashed curves show
the extrapolations ignoring the $\pi\Delta$ contributions.

The curvature in the extrapolated lowest moments of the helicity
distribution ($g_A$) increases somewhat the discrepancy between the
lattice value and experiment, although the contributions of the
$\Delta$ largely reduce this effect.  Nevertheless, there does appear
to be a residual 10--15\% underestimation of $g_A$, while at the same
time the $n=1$ and 2 moments of $\Delta u-\Delta d$ are somewhat
overestimated (although the errors on the $n=2$ lattice data in
particular are rather large).  There have been several suggestions for
the possible origin of this discrepancy, including the effects of
working with a finite volume on the lattice.  The RBCK Collaboration
has in fact observed an unusually strong dependence of $g_A$ on the
lattice volume.\cite{Sas02} This is illustrated in the top panel of
Fig.~\ref{fig:polmom} where the large open boxes and diamonds
correspond to calculations on lattices with spatial volumes of
(2.4~fm)$^3$ and (1.2~fm)$^3$, respectively.
Clearly the issue of understanding the value of $g_A$ in QCD is of
central importance in hadronic physics; moreover, resolution of this
apparent discrepancy should enable a more reliable prediction to be
made for the moments of the transversity distribution in Fig.~3,
for which there are as yet no experimental data.

\subsection{Volume dependence}
\label{sec:volume-dependence}

\begin{figure}
\begin{center}
\includegraphics[width=0.75\columnwidth]{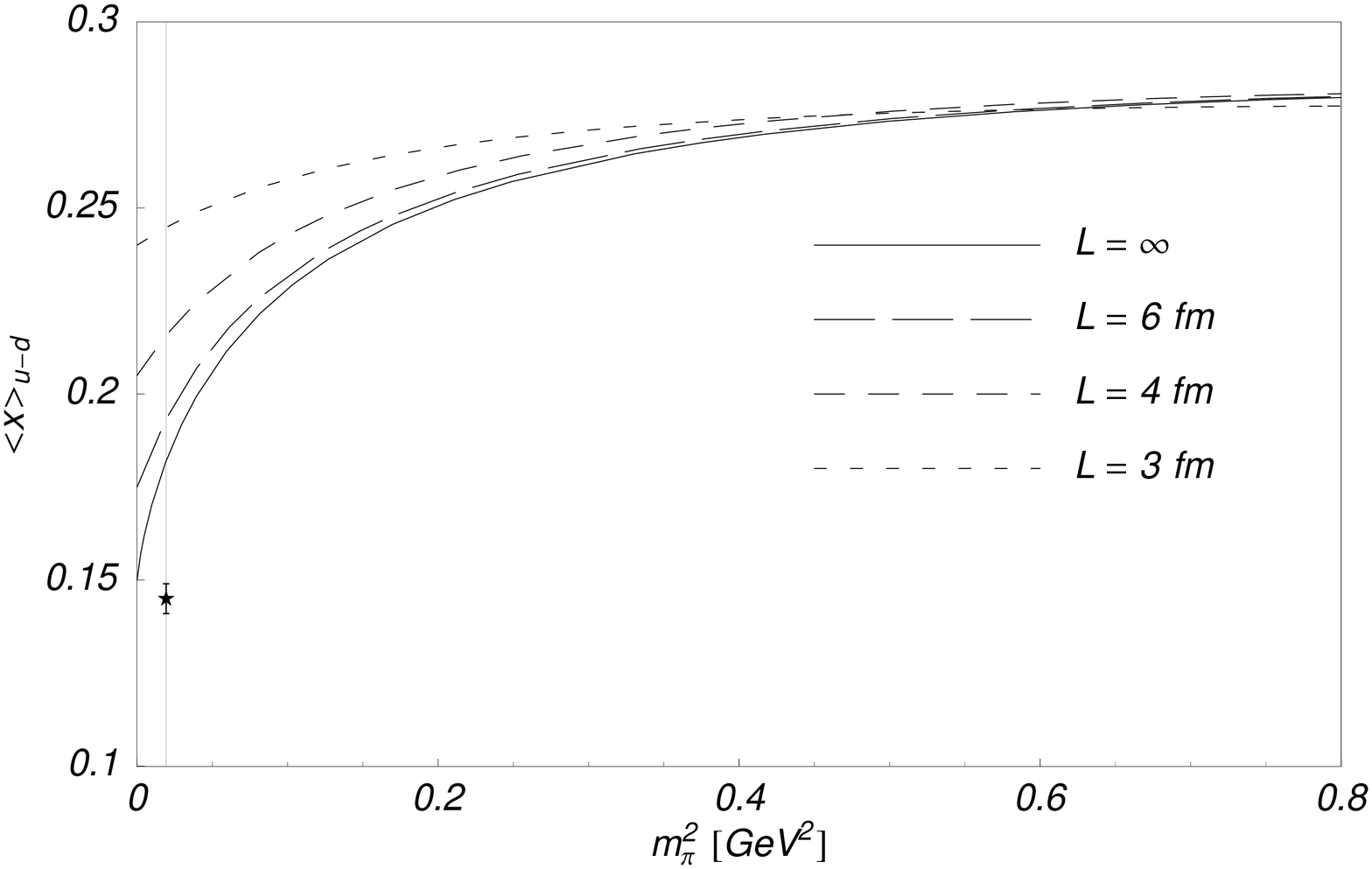}
\end{center}
\caption{Estimate of the volume dependence of the chiral curvature in
  the extrapolation of the isovector quark momentum fraction, $\langle
  x \rangle _{u - d}$, obtained with a simple cut-off on the lower
  limit of the pion three momentum, for various box side lengths $L$.}
\label{fig:V-dep}
\end{figure}

The origin of the strong lattice volume dependence of $g_A$ is likely
related to the fact that the pion field does not vanish before it
meets the lattice boundary.\cite{Jaffe:2001eb} However, there is
another important issue to consider when it comes to the curvature
that we have been discussing with regard to the chiral extrapolation
problem.  One would really like to see confirmation of this curvature
as the lattice calculations are continued to smaller pion mass and
eventually use these calculations to determine the parameter $\mu$.
This is difficult for two reasons.  First, the chiral coefficients are
smaller in quenched simulations (which are numerically less
demanding), and so the curvature may not be observable until very low
masses are reached.  Secondly, even in the case of full QCD the chiral
behavior, which is an infrared property of QCD, can be severely
altered by working on a finite volume.

For an accurate estimate of the effect of a finite lattice volume on
the chiral loops one should replace the continuum momentum integrals
in the chiral expansion by finite sums over the allowed momenta on the
lattice -- as already done for baryon and vector meson
masses.\cite{Young:2002cj} A full presentation of the results of such
an approach will be given elsewhere.\cite{Det03vol} For the present we
note that the main consequence of a finite volume is to effectively
impose a lower limit on the pion momentum integral.  One can estimate
the effect very simply by replacing the factor $m_\pi^2/(m_\pi^2 +
\mu^2)$ in Eq.~(\ref{extrap}) by $(m_\pi^2 + {\cal L}^2)/(m_\pi^2 +
\mu^2)$, where ${\cal L} = 2 \pi / Na \equiv 2 \pi / L$ is the lowest
non-zero momentum available to a pion on the lattice, with $a$ the
lattice spacing and $N$ the number of lattice sites in each spatial
dimension ({\em i.e.}, a lattice of side $L$~fm).  The results of such
a simple calculation are illustrated in Fig.~\ref{fig:V-dep}, from
which we conservatively estimate that one may need a box $L\sim 4$~fm
on a side to see substantial chiral curvature in the extrapolation,
even in full QCD.

\section{Bjorken-$x$ dependence of quark distributions}
\label{sec:bjorken-x-dependence}

In this section we present the results of recent efforts to
reconstruct the $x$ dependence of the quark distributions from their
extrapolated moments.  We shall briefly present updated results for
the unpolarized isovector distributions in the nucleon, and then focus
on new results for the longitudinally polarized valence quark
distribution.  An analogous investigation of the valence quark
distribution in the pion was performed recently,\cite{Det03pi} whilst
analysis of transversely polarized distributions awaits the
calculation of additional moments.

\subsection{Reconstruction method}
\label{sec:reconstruction}

Although the Mellin transforms,
Eqs.~(\ref{mom:unp})--(\ref{mom:trans}), have mathematically
well-defined inverses, such inversion requires knowledge of the
behavior of the moments along a contour in the complex-$n$ plane.
Clearly, evaluation of nucleon matrix elements of twist-2 operators
(on the lattice or elsewhere) cannot provide this --- through the OPE,
appropriate matrix elements determine moments only for real, integral
$n$.  In order to proceed, one must assume a parametric structure of
the underlying parton distribution.  This, however, is not a new
obstacle; the same problem is encountered in parameterizations of PDFs
from experimental data at different scales and quark distribution
functions $f(x)$ (polarized and unpolarized) are often parameterized
with the form
\begin{equation}
\label{eq:qxparam}
x f(x) = A x^{b}(1-x)^{c}(1+\epsilon\sqrt{x}+\gamma x)\, .
\end{equation}
The parameters $b$ and $c$ determine the small- and large-$x$
behavior, respectively, and have physical interpretations in terms of
Regge behavior and counting rules, whereas $\epsilon$ and $\gamma$ are
introduced to provide additional freedom.  One easily sees that the
corresponding Mellin moments are given in terms of the
$\beta$-function by
\begin{equation}
\label{eq:xnparam}
\langle x^n \rangle_f = A \left[B(1+c,b+n)
                      + \epsilon B(1+c,\frac{1}{2}+b+n)
                      + \gamma B(1+c,1+b+n) \right]\, .
\end{equation}
Given this parametric form, one can now use the various sets of
moments extrapolated from the lattice data to fit the parameters in
Eq.~(\ref{eq:xnparam}).  Since there are at most only three nontrivial
moments available for each distribution, one must first reduce the
number of free parameters by fitting $\epsilon$ and $\gamma$ to an
average of the available unpolarized\cite{UNPOLPARAM} and
polarized\cite{POLPARAM} phenomenological parameterizations. This
leaves only the parameters $A$, $b$ and $c$ to be determined from the
lattice data.
If one omits $\epsilon$ and $\gamma$, it is not clear that $b$ and $c$
retain their physical meanings since there is no longer enough freedom
in the fits for these parameters to be determined solely by the small-
and large-$x$ regions, respectively.  This is evident when one
calculates the moments of a known distribution and then attempts to
use Eq.~(\ref{eq:xnparam}) to reconstruct the distribution $f(x)$.  By
studying the dependence of the reconstructed parameters on the number
of moments, $N$, used in the fit it is apparent that in the full fit,
Eq.~(\ref{eq:xnparam}), the parameters $A$, $b$ and $c$ are almost
independent of $N$, while $\epsilon$ and $\gamma$ show some variation.
On the other hand, if $\epsilon=\gamma=0$, the remaining (purportedly
physical) parameters show significantly more variation.

As mentioned in Sec.~2 above, the different crossing symmetry
properties of the even and odd moments ({\em cf.}
Eqs.~(\ref{mom:unp})--(\ref{mom:trans})) mean that the lattice data
correspond to valence (odd under charge conjugation, $C$) moments for
$n$-even ($n$-odd) in the case of the unpolarized and transversity
(helicity) distributions, and to the total (even under $C$) moments
for the remaining cases.  Ideally, one would like several moments for
each of the $C$-odd and $C$-even distributions to reconstruct both the
valence and the total distributions (which would then allow one to
determine the sea).  In practice, not enough moments are currently
known from the lattice, and one must be content with reconstructing
the valence distributions with the help of additional phenomenology.
Namely, to recover the valence moments, one needs to shift the
relevant extrapolated lattice moments by the corresponding moments of
the sea quark distribution.  In practice, since sea quarks are
concentrated at small $x$, it is only the lowest moments ($n=0$ in the
polarized case, $n=1$ for unpolarized) that shift to any appreciable
degree.  For the unpolarized case, the non-singlet sea $\bar d-\bar u$
is well determined from Drell-Yan\cite{Towell:2001nh} and deep inelastic
data,\cite{FLAVOR_DIS} $\langle x\rangle_{\bar{d}-\bar{u}}=0.008(1)$.
On the other hand, the corresponding quantities for the polarized
distributions are poorly known and estimates need to be taken into
account in the analysis.

\subsection{Reconstructed distributions}
\label{sec:reconstr-distr}

\begin{figure}[t]
\begin{center}
\includegraphics[width=\columnwidth]{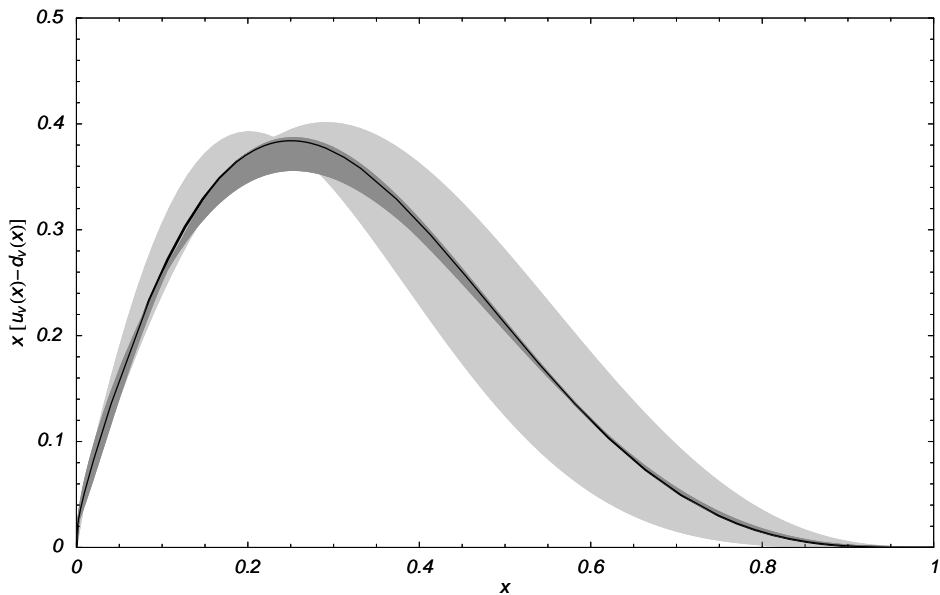}
\end{center}
\caption{Reconstructed isovector valence quark distribution 
  $x(u_v-d_v)$ in the proton at $Q^2=4$~GeV$^2$.
  The central fit curve (solid line)
  and error band (lightly shaded) are compared with the envelope of the
  phenomenological distributions\protect\cite{UNPOLPARAM} (darkly
  shaded).}
\label{fig:unpolpdf}
\end{figure}

The unpolarized valence isovector distribution
reconstructed\cite{Det01epj} from the lattice moments is shown in
Fig.~\ref{fig:unpolpdf} (solid line).  To present realistic error
bands on the reconstructed distribution, a large ensemble (of order
200 elements) of sets of moments is generated by randomly varying each
moment within the extrapolated lattice error bars.  The lightly shaded
region then corresponds to the envelope of these distributions (the
humps in the error band arise from taking the extrema of the
distributions).  The parameter values and errors are quoted in
Table~\ref{tab:partab} and are given by the mean and standard
deviation over the ensemble.  The darkly shaded region corresponds to
the average of (next-to-leading order) PDF
parameterizations\cite{UNPOLPARAM} of $u_v(x)-d_v(x)$ at $Q^2=4$~GeV$^2$.
The agreement between the reconstructed distributions and the
phenomenological parameterizations is clearly excellent.

\begin{table}[ht]
\tbl{Fit parameters for the reconstructed distributions. For each
  polarized data set, the upper (lower) rows show the parameters for
  fits with $b$ ($c$) fixed to its phenomenological value.
  In the unpolarized (polarized) fits, $\epsilon=1.96$ and $\gamma=29.1$
  ($\epsilon=-0.63$ and $\gamma=10.1$) were set to their phenomenological
  values.\protect\cite{UNPOLPARAM,POLPARAM}
\label{tab:partab}}
{\begin{tabular}{|c|ccc|}
\hline\hline
Data set	& $A$	& $b$	&	$c$		\\
\hline\hline
Unpolarized	&  0.21(5)   & -0.63(3)        & 3.8(6) \\ \hline
Polarized, set I   &  0.47(3)  & -0.51         & 2.1(2) \\
& 1.9(2)  & -0.05(7) & 3.69          			\\ \hline
Polarized, set II  & 0.28(5)    & -0.51        & 1.4(2)	\\
& 5(2) &  0.8(5) & 3.69          			\\ \hline
Polarized, set III & 0.61(4)    & -0.51        & 2.5(1)	\\
& 1.5(1)  & -0.24(6) & 3.69          			\\ \hline
\ \ \ \ Polarized, set IV\ \ \ \
& 0.70(5)    & -0.51        & 3.0(2)			\\
& 1.1(1)     & -0.39(5) & 3.69				\\ \hline\hline
\end{tabular}}
\end{table}

In contrast to the unpolarized distribution, there are a number of
issues that complicate the analysis of the polarized distributions.
Firstly, from the crossing symmetry properties of spin-dependent
structure functions, the $n=0$ moments extracted from the lattice
correspond to moments of the total distribution, $\Delta
q(x)+\Delta\bar q(x)$, and in order to construct the valence moment
one must subtract (twice) the polarized sea moment.  Unfortunately,
the only experimental information on $\langle 1 \rangle_{\Delta\bar
  u-\Delta\bar d}$ from the HERMES data\cite{Airapetian:2003ct} does
not provide strong constraints.  Consequently, we shall investigate
two scenarios: $\langle 1 \rangle_{\Delta\bar u-\Delta\bar d}=0$ (as
suggested in models in which the non-perturbative sea is generated
through meson loops\cite{THOMAS83}), and $\langle 1
\rangle_{\Delta\bar u-\Delta\bar d}=0.2(2)$ (as in quark models with
SU(6) symmetry, when the Pauli exclusion principle is
applied\cite{SCHREIBER}).

\begin{figure}[ht]
 \includegraphics[width=\columnwidth]{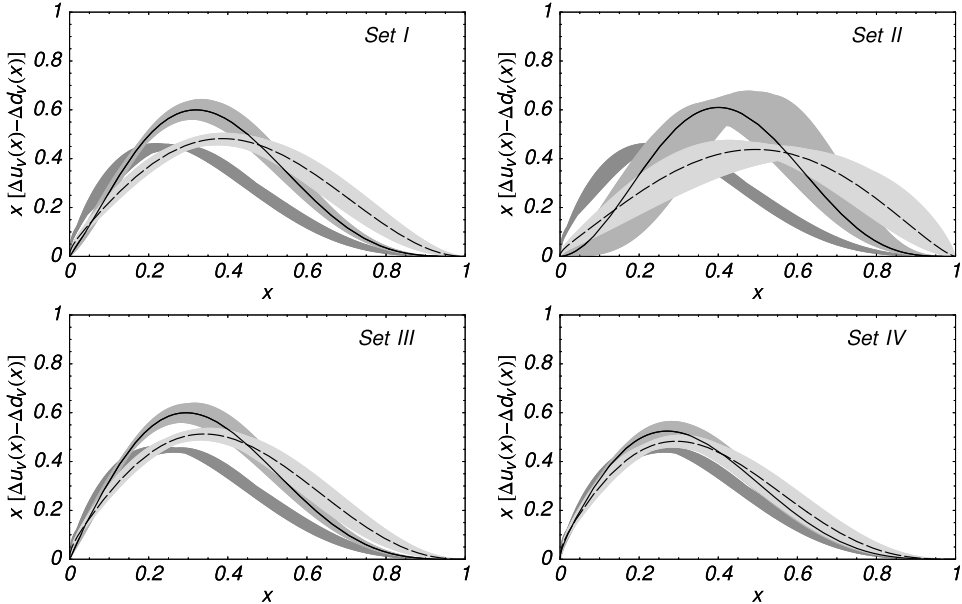}
\caption{Reconstructed isovector valence helicity distribution,
	$x (\Delta u_v-\Delta d_v)$ at $Q^2=4$~GeV$^2$.
	In each panel, the solid (dashed) curve and the
	corresponding error band is a best fit with $b$ ($c$) in
	Eq.~(\protect\ref{eq:qxparam}) fixed, while the darkly shaded
	region corresponds to the envelope of the phenomenological
        distributions.\protect\cite{POLPARAM}
	The four sets of curves are described in the text.}
\label{fig:polpdf}
\end{figure}

Secondly, as discussed above, finite volume effects have been
found\cite{Sas02} to be particularly significant in lattice
calculations of $g_A = \langle 1 \rangle_{\Delta u-\Delta d}$, and to
account for these we shall consider the effects of shifting the
extrapolated moments towards their physical value.  Finally, fitting
the three parameters $A$, $b$ and $c$ with only three moments is
unreliable (though possible), so one must presently set either $b$ or
$c$ to its phenomenological value.\cite{POLPARAM}

In Fig.~\ref{fig:polpdf} we show the reconstructed PDFs for various
shifting scenarios compared with phenomenological parameterizations.
Each panel shows the fits and error bands with $b$ fixed (solid curve,
medium shading), and with $c$ fixed (long-dashed curve, light
shading).  The darkly shaded region corresponds to the envelope of the
phenomenological distributions.\cite{POLPARAM}  In the upper left
panel, the unmodified extrapolated lattice moments are used
(data set I).
Fits to these with either $b$ or $c$ fixed give a somewhat different
reconstructed distribution, and the PDFs extracted from the lattice
are not in agreement with the phenomenological fits.

In the upper right plot we show the resulting fits that arise from
shifting the $n=0$ moment up by $0.2(2)$ to correct for the sea
contribution which enters the lattice extrapolated moment (data set
II).  The agreement between the two fits (with $b$ and $c$ held fixed)
is still poor and the errors are considerably increased.
In the lower panels, we contemplate shifts of the moments to correct
for possible finite volume (and other) effects in the lattice data and
their extrapolations.  On the left, we shift just the $n=0$ moment,
$g_A$, so that it agrees with experiment (data set III), while on the
right all three moments are shifted by the same relative percentage
towards the experimental moments (data set IV).  The latter is the
best case scenario, with the agreement between the two reconstructed
distributions significantly improved, and (not surprisingly) in
excellent agreement with the phenomenological fits.
The parameters for the various fits are given in
Table~\ref{tab:partab}.
From the variations in Fig.~\ref{fig:polpdf}, one can conclude that
the lattice data currently limit the accuracy of the reconstructed $x$
dependence of the isovector valence helicity distribution, and that it
is vital to understand the current discrepancy between the lattice and
experimental values of $g_A$.  We shall discuss the necessary
improvements in Sec.~4.

\subsection{Quark mass dependence of $x$ distributions}
\label{sec:quark-mass-depend}

In the moment analyses in Sec.~2 the most dramatic variation with
$m_\pi$ of the moments occurs at small $m_\pi$ values, as one
approaches the chiral limit.  The transition to the heavy quark limit,
on the other hand, appears to have little visible effect on the
moments, and in practice one obtains essentially identical fits to the
lattice data whether one explicitly imposes the heavy quark limit or
not.

\begin{figure}[h]
\begin{center}
 \includegraphics[width=10cm]{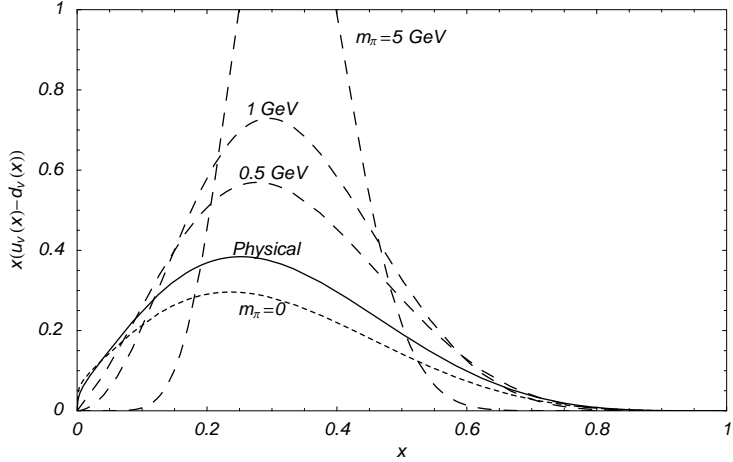}
\end{center}
\caption{Dependence of the isovector unpolarized valence quark
        distribution on the pion mass.}
\label{fig:massdep}
\end{figure}

A more graphic illustration of the change of the structure of the
nucleon from the physical region to the heavy quark limit is provided
by the quark mass dependence of the $x$ distributions.  The chiral
extrapolation formulas of Sec.~\ref{sec:extr-phys-results} provide
moments for any value of $m_\pi$, which allows one to trace how the
$x$ dependence changes as one goes from the chiral limit to the heavy
quark limit, where the distribution approaches a $\delta$-function.

As an example, we consider the unpolarized valence $u_v-d_v$
distribution, which we plot in Fig.~\ref{fig:massdep} for various pion
masses from $m_\pi=0$ to $m_\pi=5$~GeV.  As $m_\pi$ increases, the
distribution becomes more sharply peaked, with the peak moving towards
the limiting value of $x=1/3$.  This corresponds to fit parameters $b$
and $c$ becoming larger, as the $x$ dependence of the valence
distribution becomes less singular at small $x$.
The distribution at $m_\pi=5$~GeV is already rather steep, and
resembles a constituent quark--like distribution, peaking near $x \sim
1/3$.  Interestingly, the shape of this distribution is very similar
to that extracted from the moments extrapolated linearly to the
physical pion mass.\cite{Det01epj} A similar analysis can in principle
also be performed for polarized PDFs, although the present
uncertainties in the $m_\pi$ dependence of the polarized lattice
moments and the polarized sea prohibit a reliable reconstruction
of the $x$ dependence as a function of $m_\pi$.

\section{Conclusions}
\label{sec:conclusions}

The study of the PDFs of the nucleon using lattice QCD has already
achieved several notable successes. In the unpolarized case the level
of agreement between the first three non-trivial moments of the
isovector distribution and experiment is excellent. To realize this
level of agreement one must incorporate the correct leading non-analytic
behavior of chiral perturbation theory into the extrapolation of the
moments from the relatively large masses accessible in current lattice
QCD simulations to the physical regime.
Given the normalization condition and an additional three moments one
can reconstruct the $x$ dependence of the isovector valence PDF quite
reliably and, as shown in Fig.~\ref{fig:unpolpdf}, the agreement with
experiment is quite impressive.

The analysis presented here for the polarized isovector quark
distribution highlights a number of successes and also areas which
demand further study.  Amongst the successes we note:

\begin{itemize}
    \item There has been a clear demonstration of the rather unusual
  degree of sensitivity of the polarized moments to the volume of the
  lattice.  Moreover, in the case of $g_A$, the agreement with the
  experimental value definitely improves as the volume
  increases.\cite{Sas02}
  
    \item Studies of the chiral extrapolation procedure have clearly
  established the need to include virtual $\Delta$
  isobars.\cite{Det02ga,HEMMERT} Once the $\Delta$ is included there
  is a significant reduction in the chiral curvature as $m_\pi
  \rightarrow 0$, compared with that found in the unpolarized case.
\end{itemize}

For the unpolarized and especially the polarized PDFs there are a
number of challenges which must be answered in order to reach the
stage at which lattice calculations rival experimental determinations
in their accuracy. In particular:

\begin{itemize}
  
    \item In order to better constrain the chiral extrapolations
  it is important to push
  the lattice simulations to lower pion masses.  Ideally this should
  occur in full (unquenched) QCD, however, quenched, and especially
  partially-quenched simulations would also provide valuable
  information to guide the extrapolation.  For the unpolarized
  isovector case, $u-d$, one needs to confirm the predicted
  chiral curvature of the moments at low pion mass.  As suggested by
  the estimates shown in Fig.~\ref{fig:V-dep}, this will most likely
  require high statistics simulations on lattices of dimensions
  $L\sim4$~fm at a pion mass of order 300~MeV or lower.
  
    \item To better constrain the functional form of the
  $x$-dependence of the PDFs, calculations of several higher moments
  ({\em e.g.}  $n=3,\,4,\, 5$) are necessary. This is particularly
  relevant in the polarized case. Such calculations would necessitate
  the non-perturbative calculation of operator mixing coefficients.
  
    \item Finite volume effects must be explored more thoroughly and
  taken into account in future extrapolations.\cite{Det03vol} The
  effects of a finite lattice spacing ($a \ne 0$)\cite{Beane:2003xv}
  must also be incorporated.
  
    \item In the isoscalar case, one has to resolve the numerical
  problem of obtaining a reliable signal for the contribution from
  disconnected quark loops.  Furthermore, we note that separations of
  the sea quark distributions from the total isoscalar PDFs, based on
  lattice data alone, have not yet been attempted.  They will also
  require an accurate knowledge of moments with $n>3$ and evaluation
  of disconnected contributions.

\end{itemize}

With such a program we could expect significant advances in our
understanding of the quark structure of the nucleon (and other
hadrons) over the next few years. In particular, we look forward to
determinations of the various sea and transversity distributions. The
methods reviewed here will also prove useful in extracting generalized
parton distributions from recent lattice
calculations\cite{Hagler:2003jd} of non-forward matrix elements of the
various twist-2 operators.

\section*{Acknowledgments}

This work was supported by the Australian Research Council, the U.S.
Department of Energy contract \mbox{DE-FG03-97ER41014} and contract
\mbox{DE-AC05-84ER40150}, under which the Southeastern Universities
Research Association (SURA) operates the Thomas Jefferson National
Accelerator Facility (Jefferson Lab).


\end{document}